\documentclass[reprint,amsmath,amssymb,aps,superscriptaddress]{revtex4-2}
\usepackage{physics}
\usepackage{graphicx}
\usepackage[colorlinks=true,citecolor=blue,linkcolor=blue]{hyperref}

\graphicspath{{figs/}}

\def\H{\mathcal{H}}
\def\SDG{\textrm{Schr\"{o}dinger }}

\begin{document}

\title
{Optimization of conveyance of quantum particles by moving potential well}

\author{Satoshi Morita}
\email[]{smorita@keio.jp}
\affiliation{Graduate School of Science and Technology, Keio University, Yokohama, Kanagawa 223-8522, Japan}
\affiliation{Keio University Sustainable Quantum Artificial Intelligence Center (KSQAIC),
Keio University, Minato-ku, Tokyo 108-8345 Japan
}

\author{Yoshiaki Teranishi}
\affiliation{Institute of Physics, National Yan Ming Chiao Tung University, Hsinchu 30010, Taiwan}

\author{Seiji Miyashita}
\affiliation{JSR-UTokyo Collaboration Hub, CURIE, Department of Physics,
The University of Tokyo, Bunkyo-Ku, Tokyo 113-0033, Japan}

\date{\today}

\begin{abstract}
Quantum mechanical control of the position of a particle by using a trapping potential well is an important problem for the manipulation of a quantum particle.
We study the probability of successful conveyance of the particle trapped in a potential well for a given length within a given fixed time, i.e., survival probability after the motion.
For the actual motion of conveyance, we need to accelerate the particle to move and then decelerate it to stop at the destination.
Acceleration and deceleration cause a dropoff of a particle from the trapping potential well.
The relaxation of the survival probability in a process with a constant acceleration rate is studied in detail.
First, the process is studied as the relaxation of the survival probability of the trapped particle by direct numerical calculations.
The survival probability obeys an exponential decay in a long time, which is analyzed from a viewpoint of eigenvalue problem.
The value of survival probability is also estimated
by the Wentzel-Kramers-Brillouin method with connection formulas using the Airy function and the Weber function.
The value is further estimated by a method of the resonance states.
We emphasize the fact that an important source of dropoff comes from a non-analytic change of velocity at the starting point. When the rested particle begins to move, the ground state of the rest frame is redistributed to eigenstates of the moving frame, and then each eigenstate of the moving frame evolves in time. The dephasing of wave functions of the distributed populations reduces the probability of successful conveyance.
In general, a smooth start gives a small initial disturbance but it requires a large acceleration during the process to reach the destination in the fixed time which causes a larger dropoff in the process.
Considering these conflicting facts, we study the survival probability in concrete conveyance schemes, i.e., protocols with (1) a sudden change of velocity to a constant velocity, (2) a sudden change of acceleration rate to a constant acceleration, and (3) a smooth change of acceleration by studying the real-time change of populations of the adiabatic (instantaneous) eigenstates.
We observe the time evolution of the trapped probability and the population distribution during the conveyance process.
In cases that the potential well has several bound states, we propose a method to select the particle trapped at the ground state by making use of the difference of survival probabilities of bound states.
\end{abstract}

\maketitle


\section{Introduction}\label{sec:intro}
Real-time quantum dynamics is one of the most important subjects for manipulations of quantum states, and it has been studied in various topics.
For example, the Landau-Zener mechanism~\cite{Landau1932,Zener1997,Stueckelberg1932,Majorana1932} with ac field~\cite{SM-QD2011,LZac1,LZac2} and
quantum spin dynamics in metastable state under a sweeping field~\cite{SM2022} have been extensively studied.
The aspects of relaxation are also studied from the viewpoints of resonance states (non-Hermitian physics)~\cite{Siegert,Feshbach1958,Feshbach1962,Feshbach1967,Fano,Hazi-Taylor, Langer1937,Miller-Good,Connor1968,Dickinson1970,Mayer-Walter,Child-text,Hatano-nonhermite,moiseyev2011nonhermitian}.
The control of a resonance width is utilized for a scheme of laser purification in molecular vibrational cooling~\cite{atabek2013proposal,leclerc2016controlling}.
Recently, a protocol to overcome an energy barrier with a shaped pulse has been found~\cite{Miyashita2023,Miyashita2024}.
How to control a quantum state in time is a key ingredients for the quantum computing~\cite{NielsenChuang,Google2019,QuEra2022,QuEra2023,Quantinuum2021,Quantinuum2023,IBM2023}.
The time-dependent field to manipulate quantum systems in a desired way can be found by quantum optimal control theory~\cite{brif2010control}, which is originally proposed to control laser-driven dynamics of a molecule~\cite{shi1988optimal}.
Real-time dynamics also plays important roles in the quantum annealing process~\cite{Kadowaki1998,Das2008,MoritaNishimori2008,Tanaka2017,DWave2011,DWave2014}.

Among these problems of real-time dynamics, quantum effect on particle conveyance with potential trap is also an important subject.
Wu and Niu studied motion of BEC on an optical lattice~\cite{Wu-NiuNJP2003}.
Conveyance of electrons by surface acoustic waves has been discussed extensively~\cite{Tarucha2011,Byeon2021,APL119-2021}.
Coherent transport of a single electron across an array of quantum dots has been studied~\cite{Fujita2017coherent,Mills2019shuttling,Yoneda2021coherent,Tarucha2022}.
Control and visualization of atomic motion have been extensively developed by Ohmori and coworkers~\cite{Ohmori2013,Ohmori2016,Ohmori2018ACR,Ohmori2018PRL,Ohmori2020,Ohmori2022,Ohmori2023PRL,Ohmori2023arxiv}.
Quantum processors using optical tweezers to transport neutral atoms are expected to realize fault-tolerant quantum computation~\cite{QuEra2022,QuEra2023,Ohmori2022}.
Particle conveyance is also an important process in the ion-trap type quantum manipulation~\cite{Cirac1995,Quantinuum2021,Quantinuum2023}.
Theoretical study of the particle conveyance has been done in a simple case~\cite{Conveyance0}.

For the manipulation of particles trapped by a potential well, properties of dynamics in real space and real time are important.
The conveyance with a sudden change of velocity to a constant velocity was studied for a case to carry up a particle from a low-energy region to a high-energy region through a slop of potential energy~\cite{Conveyance0}. 
In the case of conveyance with a constant velocity in a flat space, the eigenstates are the same in the rest frame and the moving frame.
The amount of the escape comes only from the redistribution of populations at the initial and final points because of no acceleration during the motion.
In the case of carrying up from a low-energy region to a high-energy region, the particle can escape to the low-energy region by quantum tunneling.
That is, when the particle moves along the slope of potential energy, the particle drops off from the trapping potential with a nonzero probability given by a tunneling process. 
In a fast velocity, the escape at the starting point is large while the escape due to the tunneling is small, and the optimal condition of the velocity was studied~\cite{Conveyance0}.

In the present paper, we study the case that we move a particle from one position to another position in a fixed time by controlling motion of the position of trapping potential well.
This motion seems simple, but in practice, we need acceleration of an initially rested particle, and then deceleration to stop it at the final position.
This process requires complicated quantum operations.

For the mechanism of escape from the trapping potential, first we need to consider dropoff from the potential well at the starting point.
To begin the motion, sudden change of parameters of the Hamiltonian of the system must happen.
The set of eigenstates of the moving frame following the position of the particle is different from that of the rest frame, and the initial state which is the ground state of the rest frame is redistributed to the eigenstates of the moving frame, i.e., expressed by a linear combination of the eigenstates of the moving frame.
Due to the motions with different eigenvalues, dephasing of the redistributed wave function occurs, which causes decay of the survival probability, i.e., the probability of successful conveyance.
This type of effect owing to the initial sudden change on the survival probability was studied for a magnetic system by one of the present authors~\cite{Morita2007}.
It was shown that when the change of the position $x_0(t)$ at the initial point is proportional to $(t/\tau)^{\mu}$, i.e., $x_0(t) \propto (t/\tau)^{\mu}$, the amount of initial escape probability is proportional to $\tau^{-2\mu}$.
A change of the parameter at the final point is also important for the total escape probability.
The most abrupt start is the sudden change of velocity, which corresponds to $\mu=1$~\cite{Conveyance0}. 
To reduce the initial shock, we need to start more smoothly.
The motion of a constant acceleration rate is the next case ($\mu=2$).
The effect of acceleration is also important for the survival probability in this case.
Thus, we study this case in detail in the present paper.
As is well-known,
quantum dynamics of a system with a constant acceleration rate $a$ is equivalent to a static system in a tilted potential, in which the trapped population escapes to a low-energy region by quantum tunneling.
We obtain the energy-level diagram in the moving frame as a function of $a$. 
The $a$ dependence of eigenenergies shows a typical metastable structure which we call the metastable branch~\cite{Hatomura2016, SM2022}.
The essence of the tunneling phenomena is the dense distribution around the energy of trapped state. 

We study how the survival probability changes by investigating the time-dependent \SDG equation.
We begin by obtaining the relaxation rate purely quantum mechanically by a direct numerical method in a system with finite length.
This method is not valid after the wave reaches the boundary, but we can estimate the relaxation rate rather reasonably within the time before the wave reaches the boundary. 
To confirm the results, we also study the relaxation rate in a system with an absorbing boundary.
We find that the survival probability decays rapidly in an early stage and then relaxes exponentially at the late time.
This exponential decay of survival probability is understood as a tunneling process over a barrier in the moving frame.
We investigate this phenomenon from a viewpoint of linear combination of eigenstates of the given value of acceleration constant $a$. 
Each eigenstate oscillates with the frequency given by the eigenenergy.
To realize the exponential decay, a Lorentzian distribution is necessary. 
The dephasing of the states with this distribution gives an exponential decay in time.
Indeed, we confirm a Lorentzian distribution of eigenvalues around the metastable state.
To study this property, we analyze the process from the viewpoint of the Wentzel-Kramers-Brillouin (WKB) method, and the origin of the Lorentzian distribution of eigenvalues is confirmed. 
In the case of deep potential well, the relaxation rates obtained by the numerical method agree well with those obtained by a simple WKB method, while they do not agree with for a shallow potential as naturally expected.
This type of problem has been studied in detail as a scattering problem with a more sophisticated method~\cite{Child-text,Siegert,Feshbach1958,Feshbach1962,Feshbach1967,Fano,Hazi-Taylor,Langer1937,Miller-Good,Connor1968,Dickinson1970,Mayer-Walter}. 
For shallow cases, we applied the method using the connection relation with the Weber function, which gives better agreement with the numerical results.

The exponential-like relaxation can also be understood from the viewpoint of the resonance state in non-Hermitian dynamics for systems with only outgoing flows~\cite{Hatano-nonhermite}. We construct a corresponding non-Hermitian Hamiltonian and obtained a complex eigenvalue which gives the resonance state.
We compare thus obtained relaxation times by the above-mentioned methods and confirm the agreement among them.

Next, we study the survival probability of a particle in some concrete protocols of conveyance from a position to another position with a fixed distance ($L$) and in a time ($\tau$).
As mentioned above, the protocol of conveyance consists of acceleration and deceleration, and the acceleration rate must change in time.
Therefore, real-time control of the motion is not simple.

Concerning the effect of redistribution due to the frame change, the slower change of acceleration rate is better.
On the other hand,
to arrive at the destination in the fixed time, we need to speed up during the motion if we start smoothly.
Therefore, a large acceleration rate is necessary.
In the moving process, the eigenstates of the moving frame change according to the change of acceleration constant $a(t)$, which may cause extra escape from the trapping state during the conveyance.

Considering these competing facts, we study the dependence of the survival probability on the following three protocols of conveyance: (1) a sudden change of velocity to a constant velocity ($\mu=1$), (2) a sudden change of acceleration rate to a non-zero acceleration ($\mu=2$), and (3) a smooth change of acceleration ($\mu=3$).
Because the particle needs to stop at the final position, the integral of acceleration must be zero, and we adopt the forms $a(t)\propto \cos(\omega t)$ for the process (2) and  
$a(t)\propto \sin(2\omega t)$ for the process (3), where $\omega=\pi/\tau$.
How the survival probability changes in time as a function of $t/\tau$ during the conveyance is studied for the three cases. 
Moreover, how the populations of adiabatic eigenstates [eigenstates of the instantaneous Hamiltonian $\H(t)$] change in time is investigated, which gives information of nonadiabatic changes in the processes.

In the present paper, we mainly discuss conveyance in the system with one boundary state.
However, the cases with many bound states are also interesting.
We shortly discuss a method to control population distribution of particles in the case that the trapping potential has several bound states by making use of the difference of survival probabilities of the bound states in the protocols of conveyance.

This paper is organized as follows.
In Sec.~\ref{sec_model}, we introduce a model of the trapping potential and define the survival probability (see also Appendix~\ref{Appendix-jumpt0}).
In Sec.~\ref{sec:constant-v}, we briefly review the process with a constant velocity based on previous work~\cite{Conveyance0}.
In Sec.~\ref{sec:constant-a}, we study the case of conveyance with a constant acceleration rate in detail.
We calculate the relaxation rate of the survival probability by direct numerical simulations, the WKB method (Appendix~\ref{appendix:WKB} and \ref{appendix:Weber}), and the effective potential method for resonance states (Appendix~\ref{appendix:resonance}).
Section \ref{sec:convayance-survival} is devoted to actual conveyance processes from a position to another position within the fixed time $\tau$.
In Sec.~\ref{sec:population-change}, we discuss physical pictures of processes of dropoff from the potential well.
In Sec.~\ref{sec:multi-bound-states}, we discuss a method to control population distribution of a particle when the trapping potential has several bound states.
Lastly, we summarize the present paper in Sec.~\ref{sec:summary}.


\section{Model and method}
\label{sec_model}

In the preset paper, we study the escape of a particle from a trapping potential well in a process where we move the position of the potential well.
In this section, we introduce the model of the trapping potential and define the survival probability of the particle in the trapping potential.

\subsection{Model of trapping potential}
We consider a quantum mechanical conveyance of particles by carrying a potential well $V_\text{well}(x,t)$ in which the particles are trapped.
Here we consider a one-dimensional potential for simplicity.
The time-dependent Hamiltonian of the system is given by
\begin{equation}
{\cal H}(t) = \frac{p^2}{2m} + V_\text{well}(x,t),
\end{equation}
and the quantum state of the particle obeys the \SDG equation:
\begin{equation}
i \hbar \pdv{t} \ket{\Psi(t)}
={\cal H}(t) \ket{\Psi(t)}.
\label{eq:Psi-t}
\end{equation}
We carry the particle by shifting the potential well, i.e., $V_{\text{well}}(x, t) = V_{\text{well}}(x - x_0(t), 0)$.
The position of the potential well is controlled by a function $x_0(t)$.
Hereafter, we assume $x_0(0)=0$ and the initial Hamiltonian is denoted by $\H_0\equiv \H(0)$.

As the trapping potential well, we adopt the following form in the present paper:
\begin{equation}
V_\text{well}(x,t)=V_0 \left[
\tanh^{2}( \frac{x-x_0(t)}{w} ) - 1
\right].
\end{equation}
The eigenvalues of the bound states of $\H_0$ are given by
\begin{equation}
e_n = -\frac{\hbar^2}{2mw^2} \left(
\sqrt{\frac{2mw^2 V_0}{\hbar^2} + \frac{1}{4}} - \frac{2n+1}{2}
\right)^2,
\label{eigenvalue-pot-well}
\end{equation}
where $n$ is a non-negative integer such that the inside of the brackets in the above equation is positive~\cite{text:Landau-Lifshitz}.
The number of bound states in the trapping potential is given by this relation.
The wave function of the bound state is given by the Jacobi polynomial as
\begin{equation}
\Psi_n(x) \propto (1-\xi^2)^{\alpha/2} P_n^{(\alpha, \alpha)}(\xi),
\end{equation}
where $\xi\equiv \tanh((x-x_0(t))/w)$ and $\alpha=(-2mw^2 e_n/\hbar^2)^{1/2}$.
The special cases of the Jacobi polynomial are $P_0^{(\alpha, \alpha)}(\xi)=1$ and $P_1^{(\alpha, \alpha)}(\xi)=(1+\alpha)\xi$.

The number of bound states changes depending on the parameters, and
we control the number of bound states by tuning the parameter $m$ fixing other parameters $V_0=1$ and $w=1$.
In the present paper, we mostly study cases of one or a few bound states where quantum mechanical dynamics of the trapped state is prominent. 
The cases with many trapping states are also interesting to see how the phenomena approach the classical limit, but here we concentrate on the former cases, and leave the latter cases as a future problem.

One of the typical systems corresponding to this model is the conveyance of a neutral atom by optical traps.
For a particle with an atomic mass of $m=100 \text{ [Da]}$ (85.5~Da for Rb) and a trapping potential width of $w=1 \text{ [nm]}$,
the typical energy scale in Eq.~\eqref{eigenvalue-pot-well} is estimated as
\begin{equation}
  \frac{\hbar^2}{2mw^2} \simeq 3.3\times10^{-26} \text{ [J]}.
  \label{eq:energy-scale}
\end{equation}
If the number of bound states is small, $V_0$ and the ground-state energy have the same energy scale as Eq.~\eqref{eq:energy-scale}.
This energy scale corresponds to $2.4\text{ mK}$ and $50\text{ MHz}$ when converted to temperature and frequency.

\subsection{Transform to a moving frame}
The \SDG equation in the moving frame along a potential well is obtained by two successive unitary transformations.
First, we apply the translation operator,
\begin{equation}
U_1(t) = \exp(-\frac{ix_0(t)p}{\hbar}),
\end{equation}
which leads $U_1^{\dagger}(t)\, x\, U_1(t) = x + x_0(t)$.
The Hamiltonian is transformed as
\begin{equation}
U_1^{\dagger}(t){\cal H}U_1(t) = \frac{p^2}{2m}
+ V_\text{well}(x, 0).
\end{equation}
Using the transformation at a time $t$,
\begin{equation}
\ket{\psi(t)} \equiv U_1^{\dagger}(t) \ket{\Psi(t)},
\end{equation}
the \SDG equation is written as
\begin{equation}
\begin{split}
  &i\hbar \pdv{t}\ket{\psi(t)}\\
  &=\left[ \frac{1}{2m}\left(p-m\dot{x}_0\right)^2
  - \frac{m \dot{x}_0^2}{2} + V_\text{well}(x, 0)
  \right] \ket{\psi(t)},
\end{split}
\end{equation}
where we use Newton's notation for differentiation: $\dot{x}_0 \equiv \dv*{x_0(t)}{t}$.

Next, we shift the momentum by the unitary transformation,
\begin{equation}
U_2(t) = \exp(\frac{im\dot{x}_0 x}{\hbar}),
\end{equation}
which leads to $U_2^{\dagger}(t)\, p\, U_2(t) = p + m\dot{x}_0$.
Then, by introducing the quantum state in the moving frame,
\begin{equation}
\ket{\Phi(t)} \equiv U_2^{\dagger}(t) \ket{\psi(t)},
\end{equation}
the \SDG equation in the moving frame is written as
\begin{equation}
  \begin{split}
    &i\hbar \pdv{t} \ket{\Phi(t)} \\
    &=
    \left[
    \frac{p^2}{2m}+V_\text{well}(x, 0)
    +m\ddot{x}_0 x -\frac{m\dot{x}_0^2}{2}
    \right] \ket{\Phi(t)}.
  \end{split}
\label{SDG-eq}
\end{equation}
For the constant-velocity case [$x_0(t)= ct$], $\ddot{x}_0=0$, and $\dot{x}_0=c$ is constant, and only a constant energy shift appears.
For the constant-acceleration case [$x_0(t)= at^2/2$], $\ddot{x}_0$ is constant, and the system is equivalent to a system with additional potential $max$ added to the potential energy.
If $x_0(t)$ has higher-order dependence on $t$, the potential in Eq.~\eqref{SDG-eq} is still $t$ dependent.

\subsection{Survival probability}\label{subsec:survival-prob}

We express the quantum state in the rest frame by $|\Psi(t)\rangle$ (\ref{eq:Psi-t}), and that in the moving frame by $|\Phi(t)\rangle$ (\ref{SDG-eq}).
We define the survival probability as the overlap between the wave function at a time $t$ and the initial state in the moving frame:
\begin{equation}
p(t) \equiv \left|
\langle \Phi(0)|\Phi(t)\rangle
\right|^2.
\label{phi0phit}
\end{equation}
It is rewritten in the original frame as
\begin{equation}
  p(t) =
  \left|
  \langle \Psi(0)|U_2(0) U_2^{\dagger}(t) U_1^{\dagger}(t) |\Psi(t) \rangle
  \right|^2,
  \label{defpt-move}
\end{equation}
where we use the fact $x_0(0)=0$.

Another definition of the survival probability is given in the original frame as an overlap between $|\Psi(t)\rangle$ and the wave function given by the initial state $|\Psi(0)\rangle$ shifted to the current position of the potential well:
\begin{equation}
P(t) = \left|
\langle \Psi(0)| U_1^{\dagger}(t) |\Psi(t) \rangle
\right|^2.
\label{eq:Pt_psi}
\end{equation}
These two definitions of the survival probabilities are different, in general.
This difference comes from the difference in the momentum transformation at time $t$ and the initial time.
A sufficient condition for $p(t)=P(t)$ is that $U_2(0) U_2^{\dagger}(t)=1$, which derives $\dot{x}_0(t) = \dot{x}_0(0)$.
For the case of constant velocity sweep, the acceleration is done by a delta functionlike fashion and we need to take care of the definition of the wave function at $t=0$ carefully, which will be discussed in Appendix \ref{Appendix-jumpt0}.
In the study of the uniform-acceleration process which will be investigated in Sec.~\ref{sec:constant-a}, we have $p(t)\neq P(t)$ except $t=0$.
In the conveyance protocols which will be considered in Sec.~\ref{sec:convayance-survival}, we have $p(\tau) = P(\tau)$ at the final point of conveyance because the velocities at the initial and final times are set to be zero ($\dot{x}_0(\tau) = \dot{x}_0(0)=0$).
In the present paper, we analyze the problem in the moving frame and mainly use $p(t)$ for the survival probability.


\section{Sweep with a constant velocity}\label{sec:constant-v}
First, we review the process with a sudden change of the velocity~\cite{Conveyance0},
\begin{equation}
x_0(t) =
\begin{cases}
0, & (t<0), \\
ct, & (t \geq 0),
\end{cases}
\end{equation}
where $c$ is a constant.
In this case, the \SDG equation in the moving frame is given by
\begin{equation}
i\hbar \pdv{t} \ket{\Phi(t)} =
\left[
\frac{p^2}{2m}+V_\text{well}(x, 0)
-\frac{mc^2}{2}
\right] \ket{\Phi(t)}.
\end{equation}
Thus, the structure of eigenenergies is the same as the initial Hamiltonian $\H_0$ although the eigenenergies are shifted by a constant $-mc^2/2$, which we ignore below.

The state at time $t$ in the moving frame is
\begin{equation}
  |\Phi(t)\rangle = e^{-i\H_0 t/\hbar}|\Phi(0)\rangle
  = e^{-i\H_0 t/\hbar} U_2^{\dagger}(0)|\Psi(0)\rangle,
\end{equation}
where we use $U_1(0)=1$.
Thus, the survival probability \eqref{phi0phit} is given by
\begin{equation}
p(t) = \left| \langle \Psi(0) | U_2(0) e^{-i\H_0 t/\hbar} U_2^{\dagger}(0) |\Psi(0) \rangle \right|^2.
\end{equation}
This overlap gives the survival probability when the potential suddenly stops at a time $t$.
Expanding the initial state in the moving frame as
\begin{equation}
  |\Phi(0)\rangle = U_2^{\dagger}(0) |\Psi(0)\rangle =\sum_{k}c_k \ket{k},
\end{equation}
where $\ket{k}$ is the $k$th eigenstate of $\H_0$,
\begin{equation}
{\H}_0|k\rangle=E_k|k\rangle,
\end{equation}
we have
\begin{equation}
  \begin{split}
    p(t) =&
    \sum_k|c_k|^4 \\
    &+ 2\sum_k\sum_{k' > k}|c_k|^2|c_{k'}|^2
    \cos\left(
    (E_k-E_{k'})t/\hbar
    \right).
  \end{split}
\label{overlap-constv}
\end{equation}
The second term oscillates in time, and thus the average change of $p(t)$ is given by the first term.


\section{Uniform acceleration}
\label{sec:constant-a}

In this section, we study the escape of the particle in the case of a smooth start with a constant acceleration $a$, i.e., $x_0(t) = at^2/2$.
As mentioned in Sec.~\ref{sec:intro}, 
at the initial and finial points, the system is imposed by a nonadiabatic change of parameters in the Hamiltonian~\cite{Morita2007}.
The eigenstates vary by this change.
The ground state of the rest frame is redistributed to a linear combination of the new eigenstates of the moving frame, and the dephasing due to the difference of eigenenergies causes the escape from the trapped state.
Because the $a$ dependence of the escape rate from the trapping potential is important to understand, the mechanism of escape by quantum tunneling, we study the properties of the relaxation process in detail in this section.

\begin{figure}[t]
\centering
\includegraphics[scale=1]{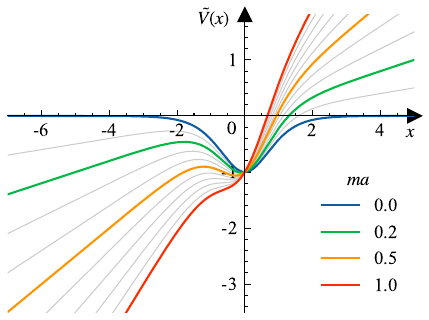}
\caption{The transformed potential energy $\tilde{V}(x)$ defined in Eq.~\eqref{eq:V_slope}.
}
\label{fig:potential}
\end{figure}

The \SDG equation in the moving frame \eqref{SDG-eq} is now
\begin{equation}
i\hbar \pdv{t} \ket{\Phi(t)} = 
\left(
\H_0 + max
\right) \ket{\Phi(t)}
\equiv \tilde{\H} \ket{\Phi(t)},
\label{eq:tilde_H}
\end{equation}
where we ignore the term of $-m(at)^2/2$ which only affects the global phase.
The present system is equivalent to the time-independent system with the transformed potential:
\begin{equation}
\tilde{V}(x) = V(x) + max.
\label{eq:V_slope}
\end{equation}
This potential is tilted by the slope $ma$ as depicted in Fig.~\ref{fig:potential}.
The second term causes a linear slope of the potential and all the eigenfunctions are extended.
Thus, in this potential, the particle is not trapped for a long time, but is temporally trapped by the local metastable potential well near $x\simeq 0$.

\subsection{Energy level diagram as a function of \texorpdfstring{$a$}{a}}

First, we study the eigenstates and eigenvalues of the Hamiltonian $\tilde{\H}$ as a function of $a$. 
Here, we adopt an approximate model that the system is confined in a region $|x|\leq L$ with a discretized mesh $\Delta x$.
The discretized time-independent \SDG equation is given by
\begin{equation}
  \tilde{E}_{k}\Phi^{(k)}_j=
  -{\hbar^2\over 2m(\Delta x)^2} \nabla^2 \Phi^{(k)}_j
  +
  \tilde{V}_j \Phi^{(k)}_j,
\end{equation}
by describing the $k$th eigenvector as $\Phi^{(k)}_j\equiv \Phi^{(k)}(j\Delta x)$, $(k=1,\cdots, N)$.
Here, $\nabla^2 \Phi^{(k)}_j \equiv \Phi^{(k)}_{j+1}-2\Phi^{(k)}_j+\Phi^{(k)}_{j-1}$ and $\tilde{V}_j=\tilde{V}(j\Delta x)$.
The eigenvalues and eigenstates are obtained by the exact diagonalization (ED) of the $N\times N$ matrix of the Hamiltonian ($N\equiv 2L/\Delta x + 1$).
In our numerical simulations, all quantities are made dimensionless, so $V_0$, $w$, and $\hbar$ are set to be unity.
Thus, the system with uniform acceleration can be specified by only two parameters, $m$ and $a$.

\begin{figure}[t]
  \centering
  \includegraphics[scale=1]{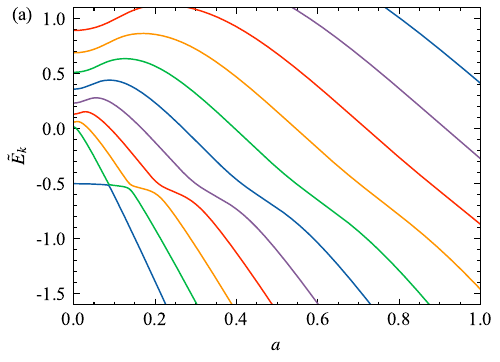}
  \includegraphics[scale=1]{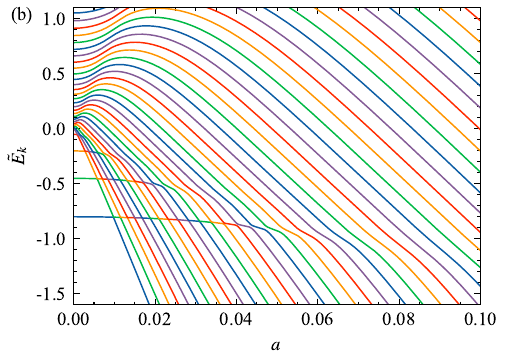}
  \caption{The energy levels as functions of acceleration $a$ for (a) $m=1$ and (b) $m=10$.
  The system is confined in a region $|x| \leq 10$ with a discretized mesh $\Delta x=0.1$.
  }
  \label{fig:energy_levels}
\end{figure}

Eigenvalues $\tilde{E}_k$ as functions of $a$ are depicted in Fig.~\ref{fig:energy_levels}.
Because the energy levels are too dense to plot for small $\Delta x$, which we used in the calculation of $p(t)$,
we show the case of $\Delta x=0.1$ and $|x|\leq 10$, which captures important characteristics of structure well.
Although there are $201$ eigenvalues for a given value of $a$, we plotted a low-energy part.
Figure \ref{fig:energy_levels}(a) is the case of shallow potential well ($m=1$).
There we find only one bound state as we know from Eq.~(\ref{eigenvalue-pot-well}).
Figure \ref{fig:energy_levels}(b) is a case of deeper potential well ($m=10$),
where we find four bound states.
In this case, the interval of energy levels of extended states become narrower because the interval between the levels is proportional to $1/\sqrt{m}$.
But, in both figures, general structure is common except the number of bound states.

At $a=0$, the eigenvalues below $0$ give bound states.
For $a>0$, the system has a linear potential, and the bound state disappears because all the states are out of the potential well at large negative $x$.
Therefore, only extended states exist (which are close to the eigenstate given by the Airy function).
Bound states in a region of small $a$ in Fig.~\ref{fig:energy_levels} are an artificial effect due to finite $L$ and this region shrinks when we increase $L$.
It should also be noted that the density of curves of eigenvalues linearly increases with $L$ and the extended states form the continuum states in the limit of $L=\infty$. 

The tracers of the bound states (nearly horizontal line starting from the bound states) 
cross other lines with avoided level crossings. 
This is a typical characteristic of the metastable state.
The gaps of avoided level crossings become larger when $a$ becomes large.
For large $a$, the potential (\ref{eq:V_slope}) does not have a metastable shape any more, and the crossing
structures are smeared around the critical value of the metastability (the spinodal point). 
This kind of sequence of avoided-level crossings is a characteristic of energy-level diagrams of the metastable states which we called the metastable branch~\cite{SM2022, Hatomura2016}.
In the case of $L\rightarrow \infty$, the energy levels become dense, but the characteristic of the diagram is maintained, which gives the so-called resonance state which we will explain later.

For a fixed value of $a$, the state is given by a linear combination of eigenstates of the Hamiltonian of $a$.
The dynamics of the state is given by evolution of each eigenstate with its eigenvalue.
In the next subsection, we study relaxation of the survival probability after a sudden change of $a$.

\subsection{Relaxation in the system with a fixed acceleration}

At a fixed value of $a$, the energy levels consist of metastablelike states (around the nearly flat lines in Fig.~\ref{fig:energy_levels}, i.e., metastable branches) and the extended states (on the lines with negative slopes).
The initial state is the ground state of the system with $a=0$.
This state is a real bound state in a trapping potential well.
At $t=0$, we suddenly change the acceleration constant ($a=0\rightarrow a)$.
Although the velocity ($\dot{x}_0=at$) is still zero at $t=0$, the eigenstates and eigenvalues in the moving frame change suddenly from those of the rest frame.

In the present case, the initial state in the moving frame is the same as the one in the original frame,
\begin{equation}
|\Phi(0)\rangle = |\Psi(0)\rangle = |0\rangle,
\end{equation}
because $\dot{x}_0(0)=0$ and $x_0(0)=0$. Let the eigenvectors and eigenvalues of $\tilde{H}$ be
\begin{equation}
\tilde{H} |\tilde{k}\rangle = \tilde{E}_k |\tilde{k}\rangle.
\end{equation}
The initial state is a superposition of eigenstates of the moving frame with different eigenenergies as
\begin{equation}
|0\rangle=\sum_{k}\tilde{d}_k|\tilde{k}\rangle,\quad \tilde{d}_k \equiv \langle{\tilde{k}}|0\rangle.
\label{dk2}
\end{equation}
Therefore, the initial state decays by dephasing owing to different eigenvalues,
that is, it evolves as
\begin{equation}
\ket{\Phi(t)} = \sum_{k} e^{-i\tilde{E}_k t/\hbar} \tilde{d}_k
\ket*{\tilde{k}}.
\label{eq:Phi_t_const_a}
\end{equation}

\begin{figure}[t]
\includegraphics[scale=1]{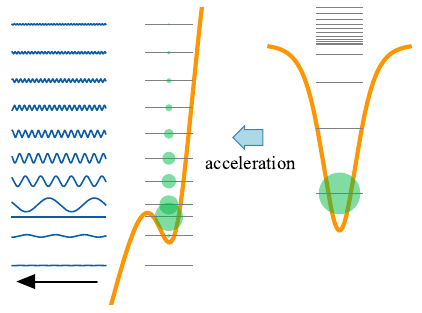}
\caption{
  Schematic picture of the dephasing caused by sudden change of acceleration.
  Right: The initial distribution is the ground state in the rest frame $|0\rangle$.
  Left: The expansion of the initial state $|0\rangle$ in the rest frame to the eigenstates of accelerated system.
  The radius of a circle is proportional to $|\tilde{d}_k|$ on the energy levels.
  The oscillation of each wave represents the probability amplitude $|\tilde{d}_k|$ and the frequency which is given by the energy difference from the level with the largest population, respectively.
}
\label{initial-disturbance}
\end{figure}

An example of this decomposition is depicted in Fig.~\ref{initial-disturbance}.
Before the acceleration, the state is the ground state of the trapping potential (right).
After the acceleration, the state is expressed by a linear combination of eigenstates of the new frame (\ref{dk2}) (left).
Each state evolves with its own frequency $e^{-i\tilde{E}_k t/\hbar}$, which causes dephasing of the states which expresses the dropoff from the trapping potential.

The survival probability is given by
\begin{equation}
  \begin{split}
    p(t)
    = &\sum_{k}|\tilde{d}_k|^4\\
    & +2\sum_{k}\sum_{k' > k}
    |\tilde{d}_k|^2|\tilde{d}_{k'}|^2
    \cos\bigl(
    (\tilde{E}_k-\tilde{E}_{k'})t/\hbar
    \bigr).
  \end{split}
\label{overlapX}
\end{equation}
The escape rate from the metastable potential well is $1-p(t)$.
We may understand the process more intuitively in the following way.
The states of high energies dephase fast and their contribution to the population in the potential well decays fast.
On the other hand, the states close to the metastable states dephase slowly and their contribution to the population in the potential well relaxes slowly as a tunneling process as depicted in Fig.~\ref{initial-disturbance}.

\begin{figure}[t]
  \centering
  \includegraphics[scale=1]{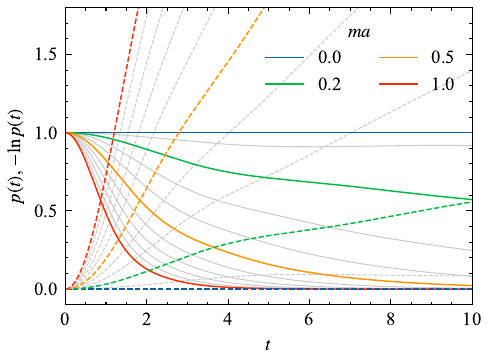}
  \caption{Decay of the overlaps $p(t)$ for $ma=0, 0.1, \dots, 1.0$ with $m=1$.
  The system is confined in $|x| \leq 200$ with a discretized mesh $\Delta x=0.1$.
  The solid (dashed) lines show $p(t)$ [$-\ln p(t)$].}
  \label{fig:overlap_p}
\end{figure}
  
We numerically obtained the relaxation of survival probability $p(t)$.
Here, we calculate $\tilde{E}_k$ and $|\tilde{k}\rangle$ by ED with the mesh of $\Delta x=0.1$.
We confirmed that the value is small enough to represent the limit $\Delta x\rightarrow 0$.
In Fig.~\ref{fig:overlap_p}, the relaxations, i.e., $\{p(t)\}$ for several values of $a$, are plotted.
There, $p(t)$ at a late stage seems an exponential decay.
To see this fact, we also plot $-\ln p(t)$.
From the slope at a late stage, we estimate the relaxation rate $\Gamma_{\rm relax}$ from the relation:
\begin{equation}
p(t) \sim e^{-\Gamma_{\rm relax} t}.
\label{prelax}
\end{equation}
The sharper slope of $-\ln p(t)$ gives a faster relaxation.
From this observation, we find that the relaxation is consistent with the above-mentioned intuitive picture.

However, $p(t)$ is given by the pure quantum dynamics given by \eqref{overlapX}, i.e., a summation of oscillating functions $e^{-i\tilde{E}_k t/\hbar}$.
The apparent exponential decay should come from the summation over many frequencies
$\{2\pi(\tilde{E}_k-\tilde{E}_{k'})/\hbar\}$
as in the case of homogeneous broadening.
For example, if the density of states $D(\omega)$ of the energy gap
$\tilde{E}_k-\tilde{E}_{k'} \equiv \hbar\omega$ has a Lorentzian form
\begin{equation}
D(\omega)={1\over a^2+\omega^2},
\label{D-omega-Lorentzian}
\end{equation}
then the dynamics is given by
\begin{equation}
\int_0^{\infty}\cos(\omega t)D(\omega)d\omega = {\pi\over 2a}e^{-at}.
\end{equation}

\begin{figure}[t]
\centering
\includegraphics[scale=1]{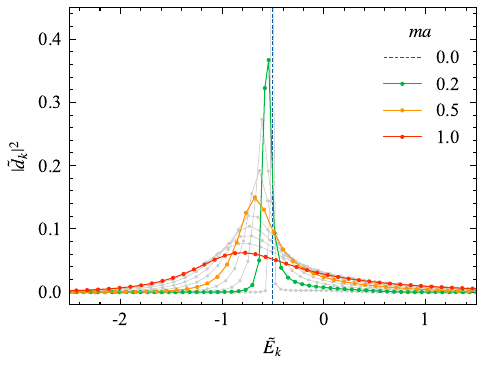}
\caption{Distribution of $|d_{\tilde{k}}|^2$ in Eqs.~\eqref{dk2} of the system---same as Fig.~\ref{fig:overlap_p}.
The dashed vertical line indicates the energy of the ground state of the initial Hamiltonian $\H_0$.
}
\label{fig:density_of_states}
\end{figure}

Although the eigenvalues are rather widely spread and the Lorentzian structure is hardly seen in Fig.~\ref{initial-disturbance}, 
we find it in the system with the larger size in which denser states exist as depicted in Fig.~\ref{fig:density_of_states}.
For $ma=0$, $\tilde{d}_k=\delta_{k,0}$ as it should.
The distribution becomes wider as $a$ increases.
The distribution locates around $\tilde{E}_k^a\simeq E_0^{a=0}$, which gives the metastable branch.

If we use a deeper potential well, there are several bound states which have different relaxation rates. In such cases, the relaxation is given by a sum of exponential decays.
At a late stage, the slowest relaxation rate dominates the relaxation.

This mechanism of relaxation can be understood by analyzing the process by the WKB method. The detailed analysis is given in Appendix \ref{appendix:WKB}.
There we construct the wave function by connecting wave functions of three regions, i.e.,
left of the barrier, inside the barrier, and right of the barrier. 
In this case, we have two turning points and use the connection formula for a single turning point twice.
This approximation should be good for high barriers, i.e., small $a$, as we confirm in Appendix \ref{appendix:WKB} Eq.~(\ref{WKB-Gamma}),
where we find Lorentzian distribution of energies around the trapped state.

In Fig.~\ref{fig:overlap_p}, however, exponential decays are observed even for large values of $a$.
This observation can be explained by using a sophisticated connection formula utilizing the asymptotic form of the parabolic cylinder equation
\begin{equation}
  \left({d^2\over dx^2}+{1\over 4}x^2-a\right)W(x)=0,
  \label{eq:parabolic}
\end{equation}
where $W(x)$ is the Weber function~\cite{Child-text}.
This method works well even for cases with low barriers.
We give a brief explanation of this method in Appendix \ref{appendix:Weber}, and give a comparison of the estimated relaxation times obtained by this method and those by the WKB method in Fig.~\ref{fig:gamma} together with values with other methods explained in the next section.

\subsection{Long-term simulations with the absorbing potential}

In the previous section, we find an exponential decay of the trapped probability $p(t)$.
However, there we confine the particle in a finite region.
Thus, when the particle reaches the boundary, it reflects and the relaxation process is disturbed.
In this sense, our simulation in the previous subsection is good only for an early stage of the relaxation.
In this subsection, we simulate the time-dependent \SDG equation with the absorbing potential which avoids the bounce of the wave packet and robustly estimate the relaxation rate $\Gamma$.

We adopt the Crank-Nicolson (CN) method to solve the time-dependent \SDG equation in the large system.
The discretized wave function is updated by the following formula:
\begin{equation}
  \begin{split}
    \frac{i\hbar}{\Delta t}(\Phi_j^{n+1} - \Phi_j^{n})
    =& -\frac{\hbar^2}{4m(\Delta x)^2}
    \left(\nabla^2 \Phi_j^{n+1} + \nabla^2 \Phi_j^{n}\right)\\
    &+ \frac{1}{2} (\tilde{V}_j^{n+1}\Phi_j^{n+1} + \tilde{V}_j^{n}\Phi_j^{n}).
  \end{split}
\end{equation}
Here, $\Phi_j^{n}\equiv \Phi(j\Delta x, n\Delta t)$,
$\nabla^2 \Phi_j^{n}\equiv \Phi_{j+1}^{n} - 2 \Phi_j^{n} + \Phi_{j-1}^{n}$,
and $\tilde{V}_j^n\equiv \tilde{V}(j\Delta x, n\Delta t)$.
If the Hamiltonian is Hermitian, the CN method conserves the norm of the wave function.
It is because this method is based on the approximation
\begin{equation}
\exp(-\frac{i}{\hbar}\tilde{\H}\Delta t) \approx
\frac{1-\frac{i}{2\hbar}\tilde{\H}\Delta t}{1+\frac{i}{2\hbar}\tilde{\H}\Delta t},
\end{equation}
and the right-hand side is still unitary.

\begin{figure}[t]
  \centering
  \includegraphics[scale=1]{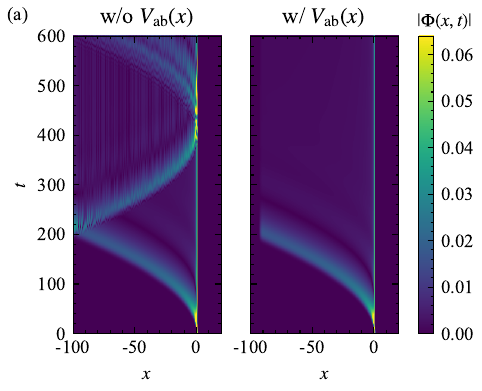}
  \includegraphics[scale=1]{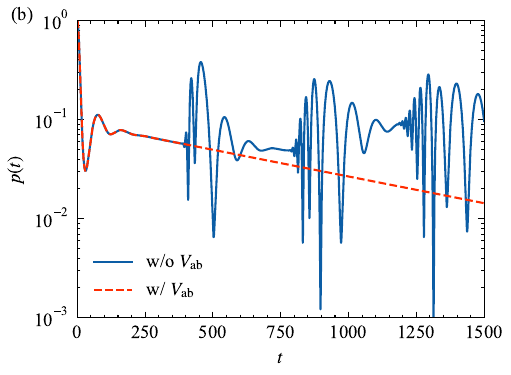}
  \caption{
    (a) The probability amplitude $|\Phi(x, t)|^2$ with (right) and without (left) the absorbing potential.
    The bounce of the wave packet at the boundary (left) is eliminated by the absorbing potential (right).
    (b) The survival probability with the absorbing potential (red dashed line) decays exponentially over a wider time range than that with the fixed boundary condition (blue curve).
    The simulations were performed with $m=100$, $ma=0.6$, $-100\leq x \leq 20$, $\Delta x=0.01$, and $\Delta t=0.1$.
    We set $w_\text{ab}=10$, $v_\text{ab}=20$ for $V_\text{ab}(x)$.
  }
  \label{fig:absorb}
\end{figure}
  
To suppress the bounce at the boundary, we add the absorbing potential~\cite{jolicard1985optical}
\begin{equation}
V_\text{ab}(x) = -i v_\text{ab} \left(
\frac{x-x_\text{min}-w_\text{ab}}{w_\text{ab}}
\right)^2
\end{equation}
for $x_\text{min}\leq x<x_\text{min}+w_\text{ab}$.
As shown in Fig.~\ref{fig:absorb}(a), 
in the simulation without $V_\text{ab}(x)$, the wave packet $|\Phi(x,t)|$ is reflected,
while the wave packet at the boundary disappears in the system with absorption.
The survival probabilities with and without the absorbing potential are shown in Fig.~\ref{fig:absorb}(b).
There, we find that exponential decay continues over a wide time range,
from which we can robustly estimate the relaxation rate $\Gamma_{\rm absorb}$,
\begin{equation}
  p(t)=p_0+Ae^{-\Gamma_{\rm absorb}t},
  \label{tau-absorb}
\end{equation}
where $p_0$ and $A$ are also fitting constants.
The $a$ dependence of the relaxation rate $\Gamma_{\rm absorb}$ is plotted in Fig.~\ref{fig:gamma} together with the relaxation rates obtained by other methods.

\subsection{Effective potential method for resonance state}\label{subsec:resonancestate}

It is known that the energy becomes complex in systems with outgoing boundary conditions~\cite{Siegert}.
For systems with only outgoing wave functions,
the conservation of probability is violated, and the Hermitian property of the system is violated.
This type of problem has been studied extensively as problems of the potential scattering, where the ideas of resonance state or the so-called non-Hermitian Hamiltonian method have been developed~\cite{Siegert,Feshbach1958,Feshbach1967,Fano,Hazi-Taylor, Langer1937,Miller-Good,Connor1968,Dickinson1970,Mayer-Walter,Child-text,Hatano-nonhermite}.

We study the present problem of decay of trapped population from the viewpoint of the resonance state.
The outgoing boundary condition is realized by an effective potential, which is defined by the Bessel function and the Hankel function in contrast to the plane wave for the flat potential.
The details are given in Appendix~\ref{appendix:resonance}. 
By this method, we obtain the relaxation rate from the imaginary part of complex energy $E$ as
\begin{equation}
  \Gamma_\text{res} = -2 \Im (E).
  \label{eq:Gamma-resonance}
\end{equation}
For $m=1$ and $a=0.2$, we obtain $\Gamma_\text{res} = 0.044288$,
which agrees well with the values obtained in the previous sections: $\Gamma_{\rm relax}=0.044290$ and $\Gamma_{\rm absorb}=0.044330$ (as plotted in Fig.~\ref{fig:gamma}).

\subsection{Comparison of relaxation constants obtained by different analyses}\label{subsec:comparison}

\begin{figure}[t]
  \centering
  \includegraphics[scale=1]{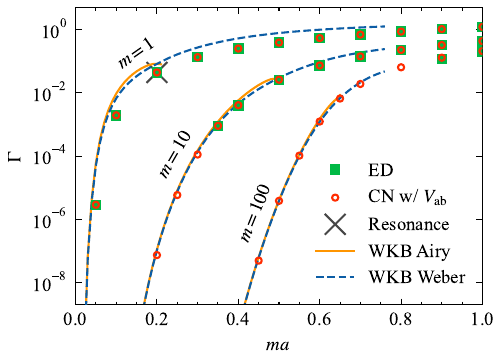}
  \caption{
  The relaxation rates $\Gamma$ as functions of $ma$ for $m=1$, $10$, and $100$ from top to bottom.
  The green squares $(\Gamma_\text{relax})$ are fitting results of $p(t)$ in Eq.~\eqref{prelax}, calculated using the exact diagonalization of the same small system as Fig.~\ref{fig:overlap_p}.
  The red circles show $\Gamma_\text{absorb}$ (\ref{tau-absorb}) by the Crank-Nicolson method with the absorbing potential.
  The large cross shows the estimation of $\Gamma_\text{res}$ by the resonance state analysis \eqref{eq:Gamma-resonance} (see also Appendix \ref{appendix:resonance}).
  The orange solid and blue dashed lines are estimated by the WKB-Airy and WKB-Weber method, respectively (see Appendixes \ref{appendix:WKB} and \ref{appendix:Weber}).
  }
  \label{fig:gamma}
\end{figure}

In the previous sections, we estimated the relaxation rate $\Gamma_{\rm relax}$ by ED \eqref{prelax} and $\Gamma_{\rm absorb}$ by simulation with the absorbing potential \eqref{tau-absorb}.
We also estimated the relaxation from the width of energy-level distribution by the WKB method (\ref{WKB-Gamma}) and by the method of the parabolic cylinder equation \eqref{Weber-Gamma}.
Moreover, we obtained the relaxation rate by the method of the resonance state (\ref{eq:Gamma-resonance}).
The results are plotted together in Fig.~\ref{fig:gamma}. 
Here, we confirm agreement among data. The values $\Gamma_\text{relax}$ and $\Gamma_{\rm absorb}$ are perfectly agree with each other, as they should be.
The WKB method gives a good estimate for the cases with large $m$ for which the potential well is deep and with small $a$ for which the slope is gentle. 
The method of the parabolic cylinder equation \eqref{Weber-Gamma} covers almost the full region of $ma$ for which the potential $\tilde{V}(x)$ has a local minimum.


\section{Optimization of conveyance}\label{sec:convayance-survival}
In the previous section, we studied the relaxation of the survival probability for the constant acceleration rate from various viewpoints, which give consistent results as they should.
In this section, we study some concrete processes $x_0(t)$ in which we carry a particle from position $0$ to $L$ in a given time $\tau$:
\begin{equation}
x_0(0)=0,\quad x_0(\tau)=L.
\end{equation}
As mentioned, for the conveyance of a particle from a position to another position, we need to accelerate the particle to move, and then decelerate it to stop at the destination.
Thus, we need to change the acceleration rate in time.
When we start with smoother acceleration, we need fast velocity in the process to arrive at the destination in the given time $\tau$.
Therefore, some optimization is necessary to realize good conveyance.

\subsection{Initial drop in the adiabatic limit}
We begin by estimating the amount of the initial drop in the adiabatic limit.
Let us consider a smooth start such as $x_0(t) \simeq at^2 / 2$ ($t\simeq 0$).
Since the process time $\tau$ gives the time scale of the system,
the acceleration rate $a$ should be scaled as $a\propto \tau^{-2}$, that is, $x_0(t)\propto (t/\tau)^2$.
In this case, according to the adiabatic theorem discussed in a previous work~\cite{Morita2007}, the drop of the survival probability at the initial time is proportional to $\tau^{-4}$ for sufficiently large $\tau$.

For a smooth start, we found the exponential decay in late time [Figs.~\ref{fig:overlap_p} and \ref{fig:absorb}(b)],
which is due to dephasing of wave functions of the dense energy levels around the trapped state forming the Lorentzian distribution.
We note that the drop $d(\tau)$ at the initial time comes from the fast relaxation of excited states with higher energies than those of the Lorentzian distribution.

For the amount of drop $d(\tau)$ at the initial time, we have an intuitive argument considering the sudden change of potential well by the change of acceleration.
Let the initial potential be $V(x)=\beta x^2$ as an approximation of the potential well around $x=0$.
Sudden change of the acceleration causes a linear potential $-\alpha x$ and the potential changes as
\begin{equation}
  \tilde{V}(x)=\beta x^2-\alpha x
  = \beta\left(x-{\alpha \over 2\beta}\right)^2-{\alpha ^2\over 4\beta}.
  \label{eq:harmonic}
\end{equation}
Here $\alpha$ is proportional to $\tau^{-2}$.
The ground states of the harmonic potential in the rest and moving frames are denoted by $|\phi_{0}\rangle$ and $|\phi_{\alpha}\rangle$, respectively.
The change between them is estimated as
\begin{equation}
  1-|\langle\phi_0|\phi_{\alpha}\rangle|^2\simeq 1-e^{\alpha^2/8\beta}\simeq {\alpha^2\over 8\beta},
  \label{eq:harmonic_approx}
\end{equation}
which corresponds to $d(\tau)\propto \tau^{-4}$.

In general, for the start of $x_0(t) \propto (t/\tau)^{\mu}$, the amount of drop $d(\tau)$ at the initial time reduces as a function of total time of process $\tau$ as $d(\tau)\propto \tau^{-2\mu}$
for sufficiently large $\tau$.


\subsection{Three typical protocols of conveyance}
Here we study three different conveyance schemes (protocols) and study how the survival probability $p(\tau)$ depends on protocols.
As concrete protocols, we adopt three protocols:
a sudden change of velocity to a constant velocity, a sudden change of acceleration rate to a non-zero acceleration, and a smooth change of acceleration:

(1) Constant-velocity protocol. The process with a constant velocity ($\mu=1$):
\begin{equation}
x_0(t)= ct,\quad c={L\over \tau}.
\label{v-const}
\end{equation}

(2) Cos-protocol. A process that begins and ends at a constant acceleration ($\mu=2$):
\begin{equation}
  a(t)=a_{1}\cos(\omega t), \quad
  x_0(t)={a_{1}\over\omega^2}\left[1-\cos(\omega t)\right].
  \label{a-cos}
\end{equation}
The coefficients $\omega$ and $a_1$ are determined by the conditions, $x_0(\tau)=L$ and $\dot{x}_0(\tau)=0$,
such as $\omega\equiv \pi/\tau$ and $a_1 = \omega^2 L/2$.

(3) Sin-protocol. A process with smoother change of acceleration ($\mu=3$):
\begin{equation}
  a(t)=a_{2}\sin(2\omega t), \quad
  x_0(t)=\frac{a_2}{2\omega}
  \left[
    t - \frac{\sin(2\omega t)}{2\omega}
  \right].
  \label{a-sin}
\end{equation}
The coefficient $a_2$ is also determined by the condition, $x_0(\tau)=L$.

\begin{figure}[t]
  \centering
  \includegraphics[scale=1]{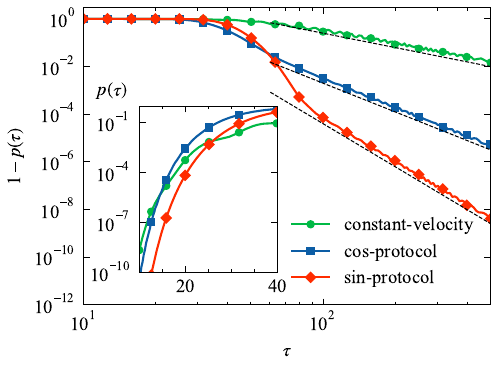}
  \caption{
  The escape probabilities $1-p(\tau)$ of three protocols, \eqref{v-const}--\eqref{a-sin}.
  The dashed lines are guides for eyes, proportional to $\tau^{-2}$, $\tau^{-4}$, and $\tau^{-6}$ from top to bottom.
  Makers are plotted at every tenth data point.
  The inset shows the survival probability $p(\tau)$ in small $\tau$ region.
  The simulations were performed in the moving frame using the Crank-Nicolson method
  with $L=50$, $m=1$, $|x| \leq 200$, $\Delta x=0.01$, and $\Delta t=0.1$.
  }
  \label{fig:conveyance_m1_excite}
\end{figure}

As discussed in Sec.~\ref{subsec:survival-prob}, we have $p(\tau)=P(\tau)$ at the end of conveyance for these three protocols.
The escape probabilities $1-p(\tau)$ are depicted in Fig.~\ref{fig:conveyance_m1_excite}.
Here, we find that as $\tau$ increases, the higher power protocol becomes better.
The dashed lines show the expected power-law dependence $\tau^{-2\mu}$ for large $\tau$.
The behavior of $1-p(\tau)$ aligns with this dependence, indicating that the drop at the start and end points dominates.
This type of dependence is similar to that of the transition probability of Landau-Zener transition in a field sweeping~\cite{Morita2007}.
On the other hand, when $\tau$ becomes smaller,
decay due to the acceleration during the conveyance begins to contribute because the value $a(t)$ becomes larger where
the decay due to the tunneling with rate $\Gamma(a)$ becomes larger.
Therefore, for intermediate $\tau$, which may happen in practical cases, the competition between the initial disturbance and the relaxation during the conveyance is serious.

\begin{figure}[t]
\centering
\includegraphics[scale=1]{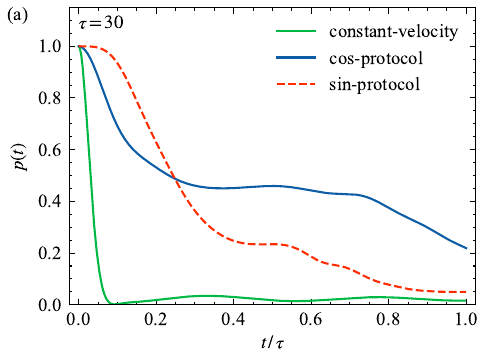}
\includegraphics[scale=1]{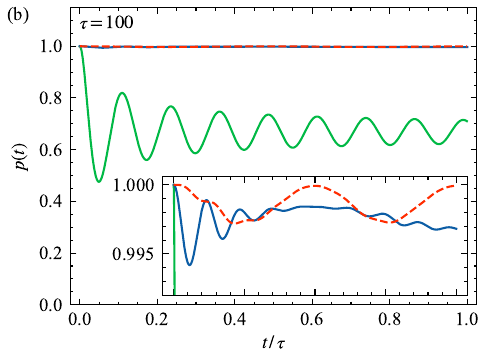}
\caption{
The survival probability $p(t)$ of (a) rapid conveyance with $\tau=30$ and (b) slow conveyance with $\tau=100$.
The inset is a magnified view around $p(t)= 1$.
The parameters are same as those used in Fig.~\ref{fig:conveyance_m1_excite}.
}
\label{fig:conveyance_m1}
\end{figure}

Figure \ref{fig:conveyance_m1} shows the time dependence of survival probabilities for the three protocols.
The survival probability of the cos-protocol (\ref{a-cos}) decreases faster than that of the sin-protocol (\ref{a-sin}) at an early stage, as we expected.
However, for the whole process,
the cos-protocol (\ref{a-cos}) gives a larger survival probability in the case of fast sweep $\tau=30$.
In the slow sweep ($\tau=100$),
the survival probability of the cos- and sin-protocols are nearly one.
However, in magnified scale, the survival probabilities show a complicated behavior, and the sin-protocol gives better conveyance.
For $\tau=100$, $\tau$ is sufficiently large to show the adiabatic behavior $1-p(\tau)\propto \tau^{-2\mu}$, and the sin-protocol (\ref{a-sin}) takes the largest value at $t=\tau$.
The long-period oscillation of $p(t)$ for the sin-protocol reflects deformation of the wave function in the trapping potential rather than the tunneling decay from the potential.
In this way, the optimization is not straightforward, but our results provide some insight into this issue.

In the constant-velocity protocol, the adiabatic eigenstates do not change in time.
The survival probability in the moving frame changes as given by (\ref{overlap-constv}), and show an oscillation mainly given by the energy gap between the bound state and the bottom of continuous states.
Thus, $p(t)$ oscillates around constants which is very small in the case $\tau=30$ and it is around $0.7$ in the case $\tau=100$.
(The precise definition of $p(t)$ in the constant-velocity protocol is given in Appendix \ref{Appendix-jumpt0}.)
Except for the very small $\tau$, the constant-velocity protocol gives a small value of $p(t)$ compared with other protocols, and thus is not considered in detail below.

\subsection{Population dynamics during conveyances}

\begin{figure}[t]
  \centering
  \includegraphics[scale=1]{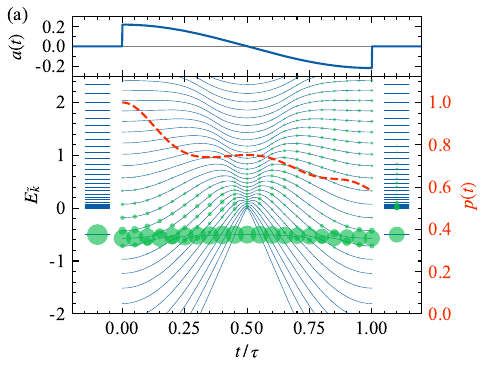}
  \includegraphics[scale=1]{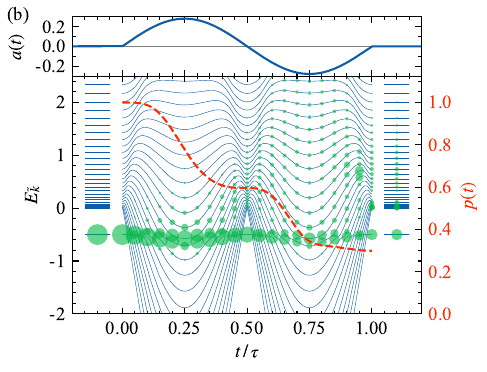}
  \caption{
  Population dynamics in fast conveyance process of (a) the cos- and (b) the sin-protocol.
  The red dashed line shows the survival probability $p(t)$.
  The thin blue lines show instantaneous eigenenergies.
  The probability amplitudes on each state are shown by the radii of the green circles.
  The horizontal blue lines shown on the left and right of the graph are the energy levels on the trapping potential before and after the conveyance.
  The change of the acceleration rate $a(t)$ is shown at top of the figure.
  We use $\tau=15$, $L=10$, $m=1$, $|x| \leq 20$, $\Delta x=0.1$,
  which exhibit $p(t)$ similar to that in Fig.~\ref{fig:conveyance_m1}(a).
  }
  \label{populations-cos}
\end{figure}

Now we study concrete examples to see how the adiabatic energy levels change and how the distribution of population changes on the levels in time.
In Fig.~\ref{populations-cos}, we depict the population dynamics for the case of the fast sweep with cos- and sin-protocols.
In general, changes of populations are not easily visible.
For example, if $L$ is large, the energy levels are too dense to see the structure.
In addition, if $\tau$ is too large, only the adiabatic process is observed; if it is too small, only the initial drop is observed.
Thus, to explain the characteristic behavior of population dynamics, we choose appropriate $L$ and $\tau$.
However, the qualitative features of the figures well reflect general features.
In the cos-protocol [Fig.~\ref{populations-cos}(a)], the population of the ground state in the initial state (the large circle at the left) scatters to excited states suddenly at $t/\tau=0$ due to the sudden change of acceleration as we discussed in Fig.~\ref{initial-disturbance}.
Then, the trapped population decreases due to the dephasing.
In the middle of the process, the value of $a$ becomes small and thus the change of $p(t)$ becomes small, and then, according to increase of $|a(t)|$, $p(t)$ decreases further.
The change of $p(t)$ is mainly determined by the instantaneous decay rate $\Gamma(a(t))$.
This type of decaying process is called adiabatic tunneling, originally proposed to explain the atomic ionization induced by intense laser fields at the low-frequency limit~\cite{Keldysh1965, PPT1966, ADK1986}.

In the sin-protocol [Fig.~\ref{populations-cos}(b)], the redistribution of population at $t/\tau=0$ is not found because the value of $a$ gradually starts from zero.
Then, the value of $a$ increases rapidly which causes redistribution of population and the adiabatic tunneling with the rate $\Gamma(a(t))$. 
In the middle of the process, $a(t)$ becomes small and, accordingly the change of $p(t)$ becomes small as well as the cos-protocol.

\begin{figure}[t]
  \centering
  \includegraphics[scale=1]{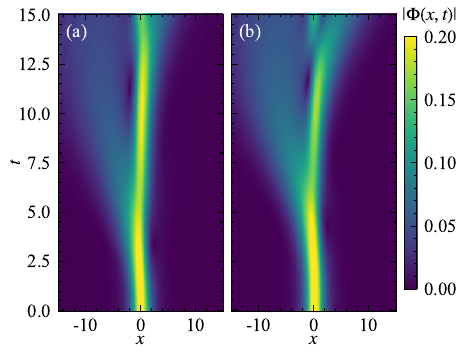}
  \includegraphics[scale=1]{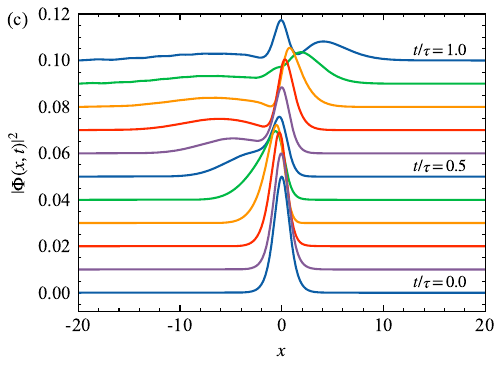}
  \caption{
    The time evolution of the probability amplitude $|\Phi(x, t)|$ in the moving frame for (a) the cos- and (b) the sin-protocol.
    (c) The time evolution of the probability distribution $|\Phi(x, t)|$ for the sin-protocol.
    The base line is shifted for each time.
    The parameters are the same as those used in Fig.~\ref{populations-cos}.
  }
  \label{conveyance_wf}
\end{figure}

The time evolution of wave functions in the moving frame for the cos- and sin-protocols are also depicted in Fig.~\ref{conveyance_wf}.
Here, we find that the weight shifts to the left during acceleration, which corresponds to the dropoff from the potential well to the negative side due the acceleration. The shift to the right is a dropoff to the positive direction due to the deceleration.

\section{Population changes with different types of acceleration schemes}
\label{sec:population-change}

In this section, we discuss the physical picture of processes of dropoff from the potential well.
We classify the changes in $a(t)$ in three ways,
which correspond to Secs.~\ref{sec:constant-v}--\ref{sec:convayance-survival}, as follows.

\begin{figure}[t]
  \centering
  \includegraphics[scale=1]{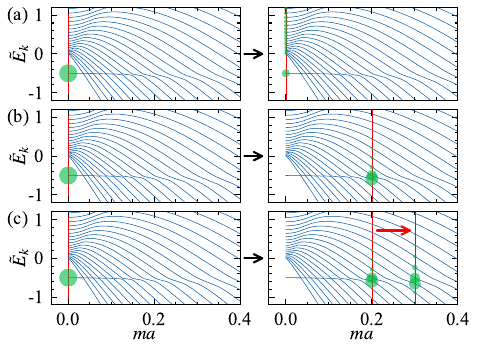}
  \caption{Sources of the reduction of survival probabilities.
  (a) If the velocity suddenly changes, the initial state is distributed into the eigenstates in the moving frame.
  (b) If the acceleration suddenly changes, the initial state is distributed into the eigenstates of the Hamiltonian with a tilted potential.
  (c) The change of the acceleration rate during the carrying process causes additional dropoff.}
  \label{fig:a-change}
\end{figure}

[A] If we change the velocity , $v(t)=0\rightarrow c$ and keep it constant $v(t)=c$,
it corresponds to the sweep with a constant velocity studied in Sec.~\ref{sec:constant-v}.
In this case,
the eigenvalue structure at $t>0$ is the same as that at $t=0$, as depicted in Fig.~\ref{fig:a-change}(a).
However, as mentioned in Sec~\ref{sec:constant-v}, eigenstates of the moving frame are different from those of the rest frame, and the initial state is distributed into eigenstates in the moving frame.
It decays 
by dephasing due to difference of eigenvalues, which gives dropoff from the potential well [see Eq.~\ref{overlap-constv}].

[B] For the case of constant acceleration $v(t)=at$ studied in Sec.~\ref{sec:constant-a},
the instantaneous Hamiltonian $\tilde{\H}$ in the moving frame is given by ${\cal H}_0-max$. Thus, in the moving frame, the eigenvalues are also different from those of $a=0$.
The $a$ dependence of eigenvalues was given in Fig.~\ref{fig:energy_levels}.
The initial state is distributed to excited states as depicted in Fig.~\ref{fig:a-change}(b),
but the adiabatic eigenstates do not change in time. 
Thus, as well as case [A], the dropoff occurs by dephasing of populations at excited states of ${\cal H}_0-max$.

In this sense, the origin of dropoff from the potential well in both cases [A] and [B] are due to dephasing.
In case [A], the bound state at $a=0$ is still a true eigenstate and isolated from other states, i.e., there is a gap between them. 
On the other hand, in case [B], there are many states around the energy of the trapped state, which is not a true eigenstate but the metastable branch. 
In the continuous limit, a continuous distribution of eigenvalues appears, and the Lorentzian distribution is formed around it as depicted in Fig.~\ref{fig:density_of_states}, which causes the exponential decay of survival probability $p(t)$ as depicted in Fig.~\ref{fig:overlap_p}.
This structure of states corresponds to the resonance state in the
treatment of non-Hermitian Hamiltonian and causes the tunneling decay
with the rate $\Gamma(a)$. In general, the localized wave function can be
decomposed into metastable resonance states with the decay rates $\Gamma_j(a)$
and rapidly decaying unstable states with high energies.
Therefore, the decay process in case
[B] can be characterized by the decay rates of resonance states $\Gamma_j(a)$
and by fast dephasings of unstable states.
If $a$ is small
in case [B], both the decaying rates $\Gamma_j(a)$ and the amount of unstable
states are reduced, giving rise to stable conveyance.

[C] In the process of time-depending acceleration $a(t)$
studied in Sec.~\ref{sec:convayance-survival},
the dropoff dynamics is characterized by the instantaneous
Hamiltonian $\tilde{\H}(t)$ in the moving frame depending on time.
As illustrated in Fig.~\ref{fig:a-change}(c), the population distribution change
in time [the populations at three different times, $a(t)=0$,
$a(t)=0.2$, and $a(t)=0.3$ are depicted], but it is
kept localized around the metastable branch, indicating that
the wave function propagates in time and follows not the adiabatic levels but
the instantaneous resonance states defined by the instantaneous acceleration
rate. The dropoff of the bound population is determined by the instantaneous
decay rates of $\Gamma(a(t))$ of the resonance state.
Thus, the process is governed by the adiabatic tunneling.
Therefore, rapid population
decay is expected when instantaneous acceleration is large. This idea explains the
results shown in Fig.~\ref{populations-cos}(b). Rapid population change is
observed around $t\simeq0.25$ and $t\simeq0.75$ at which the acceleration is maximized,
while the population is nearly stable around $t\simeq0.5$ at which the acceleration is
zero. 

Finally, we give a rough estimation of the survival probability after the conveyance using the adiabatic tunneling and the initial and final dropoff as
\begin{equation}
  p(\tau) \simeq (1-d_\text{ini}[a(t)]) e^{-\int_0^{\tau} \Gamma(a(t')) dt'} (1-d_\text{fin}[a(t)]).
  \label{eq:ptau-ad-tunnel}
\end{equation}
Here, $d_\text{ini}[a(t)]$ ($d_\text{fin}[a(t)]$) denotes the amount of the initial (final) dropoff, which is determined by the conveyance protocol.
In principle, the variational minimization of this survival probability varying the function $a(t)$ could yield the optimal protocol.
We now look at results obtained in this paper from the viewpoint of this form.
If $\tau$ is sufficiently large, the acceleration $a(t)$ during the conveyance is small, and thus the decay rate $\Gamma(a(t))$ becomes small enough to make the effect of the adiabatic tunneling negligible (Fig.~\ref{fig:gamma}).
In this case, the survival probability is governed by the initial and final dropoff,
which corresponds to the adiabatic limit, i.e., that at the large $\tau$ in Fig.~\ref{fig:conveyance_m1_excite}.
On the other hand, for small values of $\tau$, as in the inset of Fig.~\ref{fig:conveyance_m1_excite}, the survival probability is determined by the competition between the adiabatic tunneling and the initial and final dropoff.
In a similar manner, the survival probability during the conveyance process which is plotted by the red dashed curve in Fig.~\ref{populations-cos} is also estimated as,
\begin{equation}
  p(t) \simeq C e^{-\int_0^{t} \Gamma(a(t')) dt'}.
\end{equation}
Here, the coefficient $C$ takes into account all the rapid decay processes except the adiabatic tunneling.
We find that the large reduction takes place at large $|a(t)|$, which roughly corresponds to the adiabatic tunneling,
i.e., the factor $e^{-\int_0^{t} \Gamma(a(t')) dt'}$.

In the slow limit of the acceleration change, the Landau-Zener type
of nonadiabatic transition may be induced
at avoided crossings between the metastable state and extended states when the state moves along the metastable branch.
This type of transition takes place if $a(t)$ passes an avoided crossing
spending a time longer than the inverse of the energy gap of the avoided crossing.
Since the width of the Lorentz distribution of a resonance state
(shown in Fig.~\ref{fig:density_of_states}) is broader than the energy gap at the avoided crossing,
dropoff due to the dephasing of the resonance is faster than the Landau-Zener type
of nonadiabatic transition.
In other words,
to realize the adiabatic change in the Landau-Zener transition at an avoided crossing, the changing rate of $a(t)$ must be much slower,
and we conclude that it does not contribute to the dropoff mechanism in the present timescale.
Therefore, we do not analyze effects such as
the transitions
which may occur in much faster change of the acceleration rate.

\section{Conveyance by a potential well with many bound states}
\label{sec:multi-bound-states}

So far, we studied the case of a potential well with one bound state.
When we have a deeper well, there appear several bound states.
In such a case, the particle may be trapped by bound states of higher energies.
Here we propose a method to select the particles only at the ground state by tuning the carrying process.

\begin{figure}[t]
\centering
\includegraphics[scale=1]{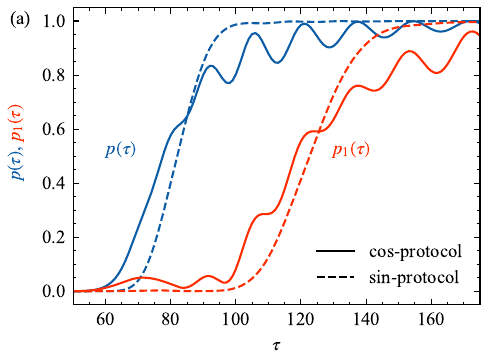}
\includegraphics[scale=1]{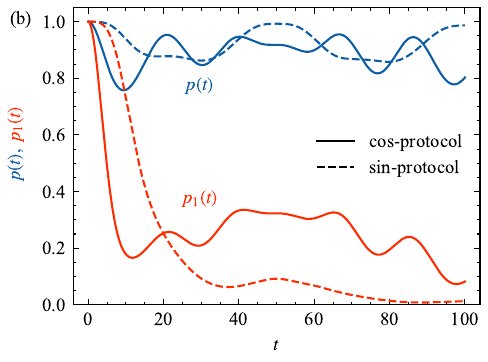}
\caption{
The survival probabilities of conveyance of the ground state and the first excited state for $m=10$.
The solid (dashed) lines use the cos- (sin-) protocol.
The parameters except $m$ are the same as those used in Fig.~\ref{fig:conveyance_m1}.
(a) $\tau$ dependence of the final survival probabilities.
(b) Time dependence of the survival probabilities for $\tau=100$.
}
\label{fig:two-particle-conveyance-m10}
\end{figure}

In Fig.~\ref{fig:two-particle-conveyance-m10}, we give an example in which the particle at the higher bound state is washed out.
Here, the survival probability is defined in the moving frame as
\begin{equation}
p_k(t) = \left|
\left\langle \Phi_k(0) \middle| \Phi_k(t) \right\rangle
\right|^2.
\end{equation}
The initial state of $| \Phi_k(t) \rangle$ is the $k$th eigenstate of the initial Hamiltonian ${\H}_0$, that is,
$\left|\Phi_k(0) \right\rangle = U_2^{\dagger}(0) |k\rangle$.
For the three protocols, \eqref{v-const}--\eqref{a-sin}, the survival probability at the final time satisfies
\begin{equation}
p_k(\tau) = P_k(\tau) \equiv \left|\langle k | U_1^{\dagger}(\tau) |\Psi_k(\tau)\rangle \right|^2,
\end{equation}
as explained in Sec.~\ref{subsec:survival-prob}.
We find that the first excited state is washed out by the conveyance process around $\tau=100$
and the sin-protocol is more effective than the cos-protocol.

In this analysis, we assume no interaction between particles, and also assume the particles are distinguishable. We hope the present analysis gives a rough idea of selections. 
However, in most plausible cases, the particles are indistinguishable.
In this case, we need the Hilbert space of squared dimensions, which will be studied in the future.

This type of dropoff of the population of high-energy bound states may be used to cool down the trapped particle. On the other hand, in cases with many bound states in the trapping potential well, the motion of conveyance causes excitations within the bound state, which is regarded as a heating effect by conveyance. Between these effects, we need another kind of optimization.

\section{Summary and Discussion}
\label{sec:summary}

We studied quantum mechanisms of drop off of particles from the potential well during a conveyance using a sweeping trapping potential.
We studied the mechanisms of exponential decay of the survival probability in a constant acceleration case in detail from various viewpoints and confirmed consistency among them.
We next studied particle conveyances with several time protocols within a fixed time.
We demonstrated three typical cases of time protocols of conveyance.
To optimize the survival probability in a fixed duration, we need a compromise between the two mechanisms, i.e., the nonadiabatic transition at the sudden changes at the start and the end of process, and decay due to the tunneling phenomena at finite $a$ in the carrying process [see Eq.~\eqref{eq:ptau-ad-tunnel}].
We also discussed the dropoff of particles from the potential well with several bound states.
The present paper should provide information not only for detailed manipulations of quantum particles, but also for general real-time procedures of quantum states.

In the present paper, we studied quantum mechanical properties of conveyance of a particle trapped by a shallow potential well, where individual motion of level dynamics plays important role.
However, in the deep potential well, there are many bound states, and the motion of state is given by an ensemble of them, which tends to the classical motion.
There, the thermalization among the states becomes of significant interest.
Such dynamics will be reported in the near future.

\section*{Acknowledgments}
We thank N.~Hatano, K.~Ohmori, and S.~Tarucha for their valuable comments.
The present paper was partially supported by JSPS KAKENHI Grants No.~JP18K03444, No.~JP20K03780 and No.~JP23H03818.
S.~Morita is supported by the Center of Innovations for Sustainable Quantum AI (JST Grant No.~JPMJPF2221).
Y.~Teranishi is partially supported by the National Science and Technology Council in Taiwan, Grant No. 109-2113-M-009-020.
The computations in this paper have been done using the facilities of the Supercomputer Center, the Institute for Solid State Physics, the University of Tokyo.

\appendix

\section{Survival probability for a constant-velocity protocol}
\label{Appendix-jumpt0}

In this appendix, we discuss the appropriate definition of survival probability in the moving frame with a constant velocity.
Let us consider a constant-velocity protocol discussed in Eq.\eqref{v-const},
\begin{equation}
x_0(t) = 
\begin{cases}
0 & (t < 0) \\
ct & (0 \leq t \leq \tau) \\
L & (t > \tau),
\end{cases}
\end{equation}
where $c \equiv L / \tau$.
The quantum state in the rest frame, $|\Psi(t)\rangle$, always satisfies the time-dependent \SDG equation in the rest frame \eqref{eq:Psi-t}.
The quantum state in the moving frame, $|\Phi(t)\rangle$ is given by the unitary transformation,
\begin{equation}
|\Phi(t)\rangle \equiv
U_2^{\dagger}(t) U_1^{\dagger}(t)
|\Psi(t)\rangle.
\end{equation}
The \SDG equation for $|\Phi(t)\rangle$ is given by Eq.~\eqref{SDG-eq}.
Since $\ddot{x}_0(t)=0$ in the constant-velocity protocol, the Hamiltonian in the moving frame is given by
\begin{equation}
\tilde{\H} = \frac{p^2}{2m} + V(x, 0),
\end{equation}
where we ignore the constant energy shift.

The sudden change of velocity causes discontinuity of the wave function in the moving frame.
Let us first consider $t=0$.
The quantum state in the rest frame should be continuous by definition.
Using the notation
\begin{equation}
|\Psi(0_-)\rangle \equiv \lim_{t \rightarrow 0-} |\Psi(t)\rangle, \quad
|\Psi(0_+)\rangle \equiv \lim_{t \rightarrow 0+} |\Psi(t)\rangle,
\end{equation}
we have
\begin{equation}
|\Psi(0_-)\rangle = 
|\Psi(0_+)\rangle = |\Phi(0)\rangle.
\end{equation}
On the other hand, since the unitary transformations around $t=0$ are given as
\begin{gather}
  U_1(0_-) = 1, \quad U_1(0_+) = \lim_{t \rightarrow 0+}
  \exp\left(-i\frac{ct}{\hbar}\hat{p}\right) = 1,\\
  U_2(0_-) = 1, \quad U_2(0_+) = \lim_{t \rightarrow 0+}
  \exp\left(i\frac{mc}{\hbar}\hat{x}\right) \equiv \tilde{U}_2,
\end{gather}
the quantum state in the moving frame is written as
\begin{gather}
  |\Phi(0_-)\rangle = |\Psi(0)\rangle,\\
  |\Phi(0_+)\rangle = U_2^{\dagger}(0_+) |\Psi(0)\rangle
  = \tilde{U}_2^{\dagger} |\Psi(0)\rangle.
  \label{Phi0+}
\end{gather}
This means that the quantum state in the moving frame is discontinuous at $t=0$.
Similarly, the quantum state in the moving frame changes discontinuously at $t=\tau$ because the velocity suddenly changes from $c$ to $0$. This change is expressed as
\begin{gather}
  |\Phi(\tau_-)\rangle = \tilde{U}_2^{\dagger}
  U_1^{\dagger}(\tau)|\Psi(\tau)\rangle, \\
  |\Phi(\tau_+)\rangle = U_1^{\dagger}(\tau)|\Psi(\tau)\rangle.
\end{gather}
For the CN method in the moving frame, the initial state should be given as $|\Phi(0_+)\rangle$ (\ref{Phi0+}). To calculate the quantum state for $t>\tau$, we need to multiply $\tilde{U}_2$ to the state at $t=\tau$ for the stopping process.

Based on these relations, let us consider the survival probability in the constant-velocity protocol.
In the rest frame, the natural definition of the survival probability is given in Eq.~\eqref{eq:Pt_psi},
\begin{equation}
P(t) \equiv |\langle \Psi(0)|U_1^{\dagger}(t) |\Psi(t)\rangle|^2.
\end{equation}
On the other hand, the survival probability in the moving frame is defined in Eq.~\eqref{phi0phit}.
However, this definition is ambiguous in the constant-velocity protocol because $|\Phi(0)\rangle$ is not well-defined.
We define two survival probabilities in the moving frame as
\begin{align}
p_-(t) &\equiv |\langle \Phi(0_-)|\Phi(t)\rangle|^2, \\
p_+(t) &\equiv |\langle \Phi(0_+)|\Phi(t)\rangle|^2.
\end{align}
It is straightforward to show that $P(t) = p_+(t) \neq p_-(t)$ for $0 < t < \tau$,
and $P(t) = p_-(t) \neq p_+(t)$ otherwise.
Therefore, we define the survival probability $p(t)$ in the moving frame for the constant-velocity protocol such that
\begin{equation}
  p(t) = 
  \begin{cases}
    p_+(t) & (0 \leq t \leq \tau) \\
    p_-(t) & (\text{otherwise}).
  \end{cases}
\end{equation}

\begin{figure}[t]
  \centering
  \includegraphics[scale=1]{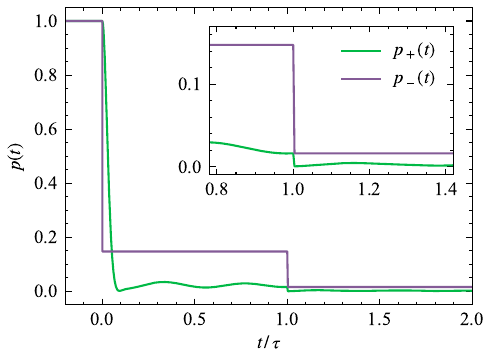}
  \caption{
    Comparison of $p_+(t)$ and $p_-(t)$ in the constant-velocity protocol.
    The inset is an enlarged view around $t=\tau$.
    The parameters are the same as those used in Fig.~\ref{fig:conveyance_m1}(a).
    }
  \label{fig:const_v_pt}
\end{figure}

We show $p_{\pm}(t)$ for the constant-velocity protocol in Fig.~\ref{fig:const_v_pt}.
While $p_-(t)$ jumps at $t=0$, $p_+(t)$ smoothly decreases from $p_+(0)=1$.
At the end of process ($t=\tau$), both the survival probabilities discontinuity changes.
As expected from the arguments above, we observe $p_+(\tau_-)=p_-(\tau_+)$.
These values are equal to the survival probability in the rest frame after conveyance, $P(\tau)$.

\section{Analysis by the WKB method} \label{appendix:WKB}

\begin{figure}[t]
  \centering
  \includegraphics[scale=1]{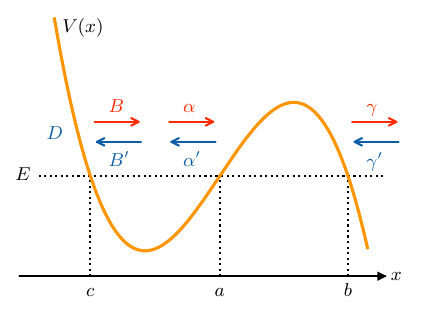}
  \caption{Typical shape of potential $V(x)$ for the resonance problem.}
  \label{WKB-metastable}
\end{figure}

Tunneling phenomena through an energy barrier is estimated by the WKB approximation~\cite{text:Landau-Lifshitz,text:Schiff}.
We study a scattering problem for the potential depicted in Fig.~\ref{WKB-metastable}.
[The direction of the outgoing wave is opposite to that in the present paper (Fig.~\ref{fig:potential}), but we use the present scheme as the usual explanation of the WKB method in textbooks.]

\begin{widetext}
Setting
\begin{equation}
  p(x)=\sqrt{2m(E-V(x))}, \quad
  \rho(x)=\sqrt{2m(V(x)-E)},
\end{equation}
wave functions in regions are
\begin{equation}
  \psi(x>b)
  ={\gamma\over \sqrt{p(x)}}\exp\left[{i\over\hbar}\int_b^x p(x')dx'\right]
  +{\gamma'\over \sqrt{p(x)}}\exp\left[-{i\over\hbar}\int_b^x p(x')dx'\right],
\end{equation}
\begin{align}
  \phi(c<x<a)
  &={\alpha\over \sqrt{p}}\exp\left[{i\over\hbar}\int_a^x p(x')dx'\right]
  +{\alpha'\over\sqrt{p}}\exp\left[-{i\over\hbar}\int_a^x p(x')dx'\right]\\
  &={B\over \sqrt{p}}\exp\left[{i\over\hbar}\int_c^x p(x')dx'\right]
  +{B'\over\sqrt{p}}\exp\left[-{i\over\hbar}\int_c^x p(x')dx'\right],
\end{align}
and for $x<c$, the wave function must be exponentially damp, and we put
\begin{equation}
\psi(x<c)=D{1\over \sqrt{\rho(x)}}\exp\left[-{1\over\hbar}\int_x^c \rho(x') dx'\right].
\end{equation}
Using the connection formula with the Airy function (see text book~\cite{text:Schiff}), 
the coefficients for $x>b$ are given by
\begin{equation}
  \begin{pmatrix}
    \gamma \\
    \gamma'
  \end{pmatrix}
  = D
  \begin{pmatrix}
    e^S + \frac{e^{-S}}{4} & -i\left(e^S - \frac{e^{-S}}{4}\right) \\
    i\left(e^S - \frac{e^{-S}}{4}\right) & e^S + \frac{e^{-S}}{4}
  \end{pmatrix}
  \begin{pmatrix}
    e^{iX} & 0 \\
    0 & e^{-iX}
  \end{pmatrix}
  \begin{pmatrix}
    e^{-i\frac{\pi}{4}} \\
    e^{i\frac{\pi}{4}}
  \end{pmatrix},
\end{equation}
where
\begin{equation}
S={1\over\hbar}\int_a^b\rho(x')dx',\quad X\equiv {1\over\hbar}\int_c^a p(x')dx'.
\label{defX}
\end{equation}
Now, introducing new variables $r$ and $\theta$ as
\begin{gather}
  r=\left[
    2e^{2S}+{e^{-2S}\over 8}+\left(2e^{2S}-{e^{-2S}\over 8}\right)\cos(2X)\right]^{1/2},\\
  \theta={\rm arg}\left({\left(e^S+{e^{-S}\over 4}\right)e^{iX}+\left(e^S-{e^{-S}\over 4}\right)e^{-iX}
  \over r}\right),
\end{gather}
\end{widetext}
we have
\begin{equation}
  \begin{pmatrix}
    \gamma \\
    \gamma'
  \end{pmatrix}
  = Dr
  \begin{pmatrix}
    e^{i(\theta - \frac{\pi}{4})} \\
    e^{-i(\theta - \frac{\pi}{4})}
  \end{pmatrix}.
\end{equation}
When $\cos(2X)=-1$,
the relation for the condition of the bound state
\begin{equation}
{1\over\hbar}\int_c^a p(x')dx'=\left(n+\frac{1}{2}\right)\pi
\label{resonance-cond}
\end{equation}
holds. In this case, $r=e^{-S}/2$, 
i.e., $\gamma$ becomes very small.
Therefore, the population on the metastable valley is large and the state represents the metastable state.
On the other hand,
when $\cos(2X)\ne -1$, $\gamma\propto e^{S}$, and thus
the population on the metastable valley is very small.

Near the energy $E_n$ which satisfies $\cos(2X)=-1$, we set
\begin{equation}
E=E_n+\Delta E,
\end{equation}
and we expand $X$ as
\begin{equation}
X(E)=\left(n + {1\over 2}\right)\pi +{\pi\over\hbar\omega}\Delta E.
\label{WKB-XE}
\end{equation}
Here, $\hbar\omega$ is a constant which corresponds to the energy gap $E_{n+1}-E_n$.
Substituting this, we have
\begin{equation}
  r(E)^2 \simeq 
  {e^{-2S}\over 4}+e^{2S}\left({2\pi\over\hbar\omega}\Delta E\right)^2.
\end{equation}
Thus, we obtain a Lorentzian form
\begin{equation}
  r(E)^{-2}={\hbar\omega\over2\pi}{\Gamma_n\over (E-E_n)^2+\Gamma_n^2}
\end{equation}
with the width
\begin{equation}
  \Gamma_n={\hbar\omega\over2\pi}e^{-2S}.
\label{WKB-Gamma}
\end{equation}
This energy distribution indicates an exponential decay of the initial wave function in time in pure quantum dynamics.
The results are plotted in Fig.~\ref{fig:gamma}.

Using an outgoing boundary condition~\cite{Siegert}, it is known that the energy becomes complex.
When we assume $\gamma'=0$, i.e., only an outgoing wave exists, the following relation holds:
\begin{equation}
e^S\cos(X)-i{e^{-S}\over 4}\sin(X)=0.
\label{WKB-gamma0}
\end{equation}
This equation does not have a real solution of the energy.
However, substituting (\ref{WKB-XE}) to (\ref{WKB-gamma0}), we have the imaginary part of the complex energy:
\begin{equation}
  \Delta E \simeq -i\frac{\hbar\omega}{4\pi}e^{-2S}.
\end{equation}
Comparing it with $E=E_n-i\Gamma_n/2$ also yields the decay rate (\ref{WKB-Gamma}).
This state with the imaginary energy is called the resonance state.

\section{The method using the connection relation by the parabolic cylinder equation}\label{appendix:Weber}

The above discussion assumes that the three classical turning points in Fig. 15 are well separated and the connection formula for an isolated turning point using the Airy function is applicable for each of them. This assumption, however, breaks for a large acceleration case, where two turning points $a$ and $b$ become close to each other. In such cases, we can utilize the connection formula based on the parabolic cylinder function to obtain the relations between $(\alpha, \alpha^\prime)$ and $(\gamma, \gamma^\prime)$, namely~\cite{Child-text},
\begin{equation}
  \begin{pmatrix}
    \gamma \\ \gamma'
  \end{pmatrix}
  =
  \begin{pmatrix}
    (1+\kappa^2)^{1/2}e^{-i\phi} & -i\kappa \\
    i\kappa & (1+\kappa^2)^{1/2}e^{i\phi}
  \end{pmatrix}
  \begin{pmatrix}
    \alpha \\ \alpha'
  \end{pmatrix},
  \label{weber-connection}
\end{equation}
where
\begin{equation}
\kappa = e^{-S},
\end{equation}
and
\begin{equation}
\phi=\arg\Gamma\left(\frac{1}{2}+i\frac{S}{\pi}\right)
-\frac{S}{\pi}\ln\left|\frac{S}{\pi}\right|+\frac{S}{\pi}.
\end{equation}
Here $\Gamma$ is the gamma function. The connection formula 
Eq.(\ref{weber-connection}) is exact for the quadratic potential barrier, and it works as a uniform approximation for general potential barrier problems with two turning points, including asymmetric potential barriers. 

The resonance width $\Gamma_n$ obtained with the use of 
Eq. (\ref{weber-connection}) is given by
\begin{equation}
\Gamma_n={2\hbar\omega\over\pi} \frac{(1+\kappa^2)^{1/2}-\kappa}{(1+\kappa^2)^{1/2}+\kappa},
\label{Weber-Gamma}
\end{equation}
Here $\omega$ is obtained from Eq.(\ref{resonance-cond}), with the modified definition 
of $X$ given by
\begin{equation}
X\equiv {1\over\hbar}\int_c^a p(x')dx' -\frac{\phi}{2},
\end{equation}
where $\phi$ is the phase shift due to the effect of two turning points. 

Here we compare the relaxation rates obtained by the connection formula using the Airy function [Eq.~(\ref{WKB-Gamma})],
and using the parabolic cylinder function [Eq.~(\ref{Weber-Gamma})]. As is shown in this figure, the formula with the Airy function overestimates the relaxation rates when the energy is close to the potential barrier maximum. It should be noted that, in this method, we do not worry about the breakdown of the connection formula of the Airy function even when the two turning points $a$ and $c$ come close together. This is because we are interested in the resonance case defined by Eq.~(\ref{resonance-cond}), which guarantees that the two turning points are not close together. 

\section{Imaginary eigenvalue of the resonance state}\label{appendix:resonance}

\begin{figure*}[t]
  \centering
  \includegraphics[scale=1]{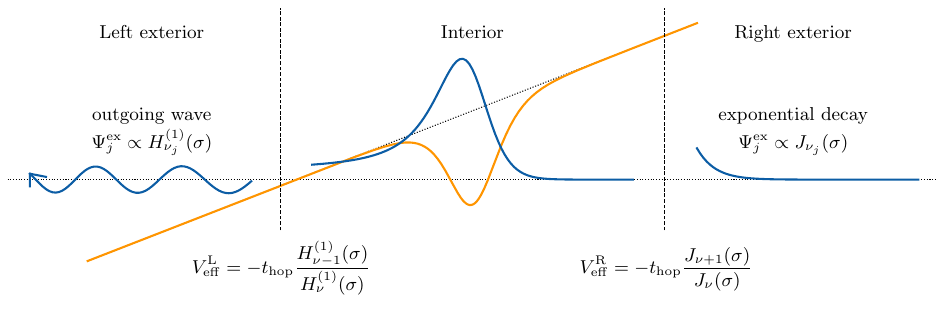}
  \caption{Connection of the wave function to out-going wave functions at boundaries.
  }
  \label{fig:resonance-connection}
\end{figure*}

In this appendix, we study the present model from the viewpoint of resonance state.
To relate the decay of population to the resonance state which is an eigenstate of non-Hermitian Hamiltonian representing the outgoing boundary condition~\cite{Siegert},
we modify
the time-independent \SDG equation in the discretized space.
It is given by
\begin{equation}
-t_\text{hop} (\Phi_{j+1} - 2\Phi_j + \Phi_{j-1}) + V_j \Phi_j = E \Phi_j,
\end{equation}
where $t_\text{hop}\equiv \hbar^2/2m(\Delta x)^2$, $\Phi_j=\Phi(j \Delta x)$, and $V_j=\tilde{V}(j \Delta x)$.
To express outgoing boundary condition, we set the following condition.
First, we separate the system into two parts, the interior ($|j|\leq L$) and exterior.
The boundary $|j|=L$ is chosen such that the potential well is sufficiently small in the exterior.
For an initial guess of the complex eigenvalue $E$, we can calculate the exact solution $\Phi_j^\text{ex}(E)$ of the exterior problem for $|j|\geq L$.
We define the effective potential at the right boundary ($j=L$) as
\begin{equation}
V_\text{eff}^\text{R}(E) \equiv -t_\text{hop}
\frac{\Phi_{L+1}^\text{ex}(E)}{\Phi_{L}^\text{ex}(E)}.
\end{equation}
Then, the \SDG equation at the right boundary is transformed as
\begin{equation}
-t_\text{hop} (- 2\Phi_L + \Phi_{L-1}) + (V_L + V_\text{eff}^\text{R}(E)) \Phi_L = E \Phi_L.
\label{resonance-SDGE}
\end{equation}
The same transformation applies to the left boundary ($j=-L$) with the effective potential $V_\text{eff}^\text{L}(E) \equiv -t_\text{hop}
\Phi_{-L-1}^\text{ex}(E) / \Phi_{-L}^\text{ex}(E)$.
Since the interior \SDG equation has been rewritten in a closed form, we can numerically diagonalize it.
The resulting complex eigenvalues $e_k(E)$ depend on the initial guess $E$, and its imaginary part is negative under the outgoing boundary condition.
The most stable resonance state has the largest (i.e., the smallest absolute value) imaginary part of $e_k(E)$.
Finally, we solve the self-consistent equation $e_k(E)=E$ by minimizing $|e_k(E)-E|$ and obtain the complex eigenenergy of the resonance state.
  
The choice of the exact solution in the exterior region $\Phi_{L+1}^\text{ex}(E)$ depends on the model.
In the present model, the discretized \SDG equation with $V_j = (ma\Delta x) j$ is expressed by the recurrence relation of the Bessel function. 
The appropriate choice of the exact solution $\Phi_{j}^\text{ex}(E)$ in the exterior is depicted in Fig.~\ref{fig:resonance-connection}.
That is, we choose the Bessel function of the first kind,
\begin{equation}
\Phi_{j\geq L}^\text{ex}(E)\propto J_{\nu_j}(\sigma),
\end{equation}
for the right exterior and the Hankel function of the first kind,
\begin{equation}
\Phi_{j\leq -L}^\text{ex}(E)\propto H_{\nu_j}^{(1)}(\sigma),
\end{equation}
for the left exterior, where
\begin{equation}
\nu_j=j+\sigma\left(1-{E\over 2t_\text{hop}}\right),
\quad \sigma={2t_\text{hop}\over ma\Delta x}.
\end{equation}

\begin{figure}[t]
  \centering
  \includegraphics[scale=1]{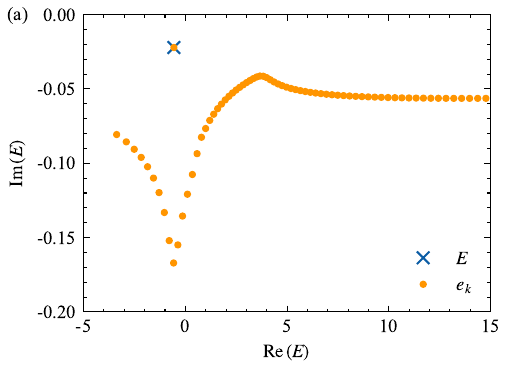}
  \includegraphics[scale=1]{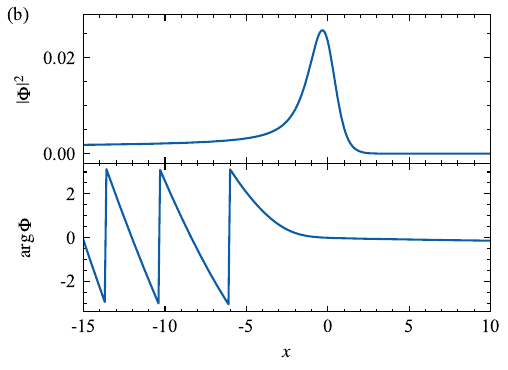}
  \caption{(a) Eigenvalues of the non-Hermitian Hamiltonian of the present model in the complex plane.
  The resonance state is the closest to the real axis and its eigenvalue satisfies the self-consistent equation.
  (b) The eigenstate of the resonance state.
  The simulation was performed with $m=1$, $ma=0.2$, $-20\leq x \leq 20$, and $\Delta x = 0.1$.}
  \label{fig:resonance-solution}
\end{figure}

The eigenfunction for the resonance model is depicted in Fig.~\ref{fig:resonance-solution}(b).
In the upper panel, the absolute value $|\Phi|^2$ gives the profile of trapped components.
As the time goes, the profile does not change, but the amplitude is reduced exponentially with the decay rate $\Gamma_\text{res}$.
The long tail in $x<0$ represents the component that escapes through the potential barrier by tunneling effect.
On the other hand, in $x>0$, $|\Phi|^2$ dumps rapidly because of the linear potential due to acceleration.
In the lower panel, the phase parts is depicted.
The phase $\arg(\Phi)$ increases when $x$ goes negative, which indicates a flow to the left.
This observation is consistent with the out-going boundary condition.

We obtain the self-consistent solution for $m=1$, $ma=0.2$,
\begin{equation}
E=-0.56123- i \times 0.022144,
\label{eq:Eresonance}
\end{equation}
by solving the minimization problem using the Powell method.
The eigenvalue distribution in the complex energy plane is depicted in Fig.~\ref{fig:resonance-solution}(a).
The eigenvalue of the resonance state is the closest to the real axis and isolated from faster relaxation modes as depicted in Fig.~\ref{fig:resonance-solution}(a).
The imaginary part gives the decay of population as
\begin{equation}
\Gamma_\text{res}=-2\Im(E)=0.044288.
\end{equation}

\bibliography{Conveyance2}

\end{document}